\documentclass[12pt]{article}
\usepackage{amssymb}

\usepackage{latexsym}
\usepackage{amsmath}

\begin{document}

\title{Electromagnetic Dipole Radiation Fields, Shear-Free Congruences and Complex
Center of Charge World Lines}
\author{Carlos Kozameh\thanks{%
kozameh@famaf.unc.edu.ar},  \thinspace Ezra Newman\thanks{%
newman@pitt.edu} \thinspace  \\
%EndAName
{\footnotesize {\ }}FaMAF, Universidad Nacional de C\'{o}rdoba, 5000
C\'{o}rdoba, Argentina,{\footnotesize {\textit{\ }}}\\
Dept of Physics and Astronomy, Univ. of Pittsburgh. Pgh PA.15260}
\date{4.2.2005}
\maketitle

\begin{abstract}
We show that for asymptotically vanishing Maxwell fields in Minkowski space with non-vanishing total charge, one
can find a unique geometric structure, a null direction field, at null infinity..  From this structure a unique
complex analytic world-line in complex Minkowski space that can be found and then identified as the
\textit{complex center of charge}. By ''sitting'' - in an imaginary sense,  on this world-line both the
(intrinsic) electric and magnetic dipole moments vanish. The (intrinsic) magnetic dipole moment is (in some
sense) obtained from the `distance' the complex the world line is from the real space (times the charge). This
point of view unifies the asymptotic treatment of the dipole moments. For electromagnetic fields with vanishing
magnetic dipole moments the world line is real and defines the real (ordinary center of charge). We illustrate
these ideas with the Lienard-Wiechert Maxwell field. In the conclusion we discuss its generalization to general
relativity where the complex center of charge world-line has its analogue in a complex center of mass allowing a
definition of the spin and orbital angular momentum - the analogues of the magnetic and electric dipole moments.
\end{abstract}

\pagenumbering{arabic}

\section{Introduction}

The relationship between the asymptotic electromagnetic fields seen by a
distant observer and the multipole moments (both electric and magnetic) of
the source producing this field is a problem worth reconsidering for several
reasons. One reason for our interest lies in the ambiguity in answering the
question: what are the values of the moments at any given time? If we are
looking at and calculating the moments directly from integrals over the
sources, their values will depend on the choice of the space-like 3-surfaces
used for their evaluation. If we attempt to evaluate the moments, via the
behavior of the asymptotic fields, the values again depend on the choice of
two-surface integrals at null infinity ($\frak{I}$) for their evaluation.
Another, and perhaps more important, reason for our interest is that in
general relativity (GR) a similar - but far more difficult problem arises;
how does one even define at null infinity the orbital and spin angular
momenta. In this case the ambiguities have been considerable and a general
consensus does not exists.

In particular, here we will be interested in studying only the electric and
magnetic dipole moments\thinspace for an arbitrary asymptotically vanishing
(real) Maxwell field. We will not be concerned with the real interior dipole
sources of the field but only what can be seen or perceived from studying
the fields at infinity. The relationship, described here, is between these
asymptotic Maxwell fields and the (perceived from infinity) motion of
certain fictional sources that are generating the dipole fields. In
addition, this analysis serves as a preparation for the much more difficult
issue of defining center of mass motion and the related orbital and
spin-angular momentum in general relativity.

The point of view developed and promulgated here, though it is completely
based on the ordinary Maxwell equations in Minkowski space, is far from
conventional. Using the existing asymptotic structure of any asymptotically
vanishing Maxwell field (always assuming a non-vanishing total charge), will
find a unique geometric structure from which there is associated a \textit{%
unique }complex world-line on complex Minkowski space. \textit{We} \textit{%
identify and refer to this world-line as the complex center of charge
world-line}.

For this arbitrary Maxwell field, roughly, we will \textit{interpret }the
dipole moments as if they were determined by the 'distance' between the
complex world-line and the coordinate origin. Again roughly speaking, the
real part of this \textit{complex center of charge world-line} (times the
total charge) defines \textit{the `intrinsic' electric dipole moment}
whereas the imaginary part (times the charge) defines the `\textit{
intrinsic' magnetic dipole moment}. We thus have an `\textit{intrinsic'
complex dipole moment. }In some sense - to be described in detail - this%
\textit{\ `intrinsic' complex dipole moment }acts as the generating function
for various standard values of the dipole moment. The basic observation that
lies behind the use of the complex world-line is that it came directly from
the following \textit{geometric structure on }$\frak{I}$: From any given
complex world-line, we obtain a specific (asymptotically shear-free) \textit{%
real null direction field} on $\frak{I}$ and the converse, any
(asymptotically shear-free) \textit{real null direction field} on $\frak{I}$
defines a complex world-line in complex Minkowski space. By certain weighted
integrals of this direction field, taken over any 2-sphere, i.e., cuts of $%
\frak{I}$, we obtain the standard asymptotic definitions of the electric and
magnetic dipole moments. We point out that if the world-line were \textit{%
real} then the field would not possess a magnetic dipole moment and the
world-line would coincide with the ordinary (real) center of charge.

We want to emphasize that in the use of these ``complex world-lines'', we
are not suggesting that these complex world-lines should be treated as if
they were \textit{`real'} even though an idealized observational
prescription for `seeing them' does appear to exist. At the moment they are
just an excellent bookkeeping and unifying device. Any possible deeper
significance will have to wait for future developments.

In addition, we claim that this observation concerning the complex
world-line and the null direction field leads to an understanding or
explanation of the Schwarzschild to Kerr and the Reissner-Nordstrom to
charged Kerr transformation 'trick' of several years ago.\cite{gyro}

This work is divided as follows. In Section 2 we present some geometrical
formulae concerning the properties of $\frak{I}$, i.e., of future null
infinity, that will be used throughout this work. In Section 3 we study, as
an example, the Lienard-Wiechert (LW) Maxwell fields. The real world-line of
a charged point particle, that is the source of this LW field, coincides
with the \textit{center of charge} discussed earlier.. In this simple case
the \textit{null direction field alluded to (above)} are the surface forming
null geodesics formed by the light-cones of the source. In Section 4 we
proceed to general asymptotically vanishing Maxwell fields and show how the
unique \textit{complex center of charge world-line} is found. In the general
case, when magnetic dipole radiation is present, the null geodesic field has
twist. In the conclusion, we summarize the ideas presented and describe how
these ideas will be applied to GR.

\section{Null Tetrads at $\frak{I}$}

The main results derived in this work will arise from the use of two different null tetrads defined in the
neighborhood of future null infinity $\frak{I}$. It is thus worth outlining the different geometrical
constructions associated with these tetrads. First a word of caution: almost all the calculations that we
perform are done at $\frak{I}$ and therefore the use of the conformal picture is, in some sense, most
appropriate. There is, however, a danger from the ambiguity in the choice of conformal factor on $\frak{I} $ so
that we feel it is safer and clearer to use the physical space-time picture in the limit as we approach
$\frak{I}$.

We start with standard Bondi coordinates on null infinity $(u_{B},\zeta ,\overline{\zeta })$ obtained from the
intersection of the light-cones from a time-like geodesic with $\frak{I}$. The intersection of these cones is
given by $u_{B}=const.$ The generators of the null geodesics on $\frak{I}$ are labelled by ($\zeta
,\overline{\zeta }).$ Associated with this coordinate
system we introduce a null vector, $n_{B}^{a},$ along the generators of $%
\frak{I}$ and complex null vectors, $m_{B}^{a},$ $\overline{m}_{B}^{a},$
tangent to the unit sphere. The tangent vectors to the Bondi light-cones
define surface-forming null vectors, $l_{B}^{a}$ , orthogonal to the $%
u_{B}=const$ cuts of $\frak{I}$. In this way we gather (with $r$ an affine
parameter along the null geodesics) a coordinate system $(r,u_{B},\zeta ,%
\overline{\zeta })$ together with a Bondi tetrad $%
(l_{B}^{a},n_{B}^{a},m_{B}^{a},\overline{m}_{B}^{a})$ in a neighborhood of $%
\frak{I}$, i.e., $r\sim \infty $.

For completeness, we give the expressions for the tetrad vectors in
Minkowski coordinates:
\begin{eqnarray}
l_{B}^{a} &=&\frac{\sqrt{2}}{2}(1,\frac{\zeta +\overline{\zeta }}{1+\zeta
\overline{\zeta }},-i\frac{\zeta -\overline{\zeta }}{1+\zeta \overline{\zeta
}},\frac{-1+\zeta \overline{\zeta }}{1+\zeta \overline{\zeta }});
\label{tetrad} \\
m_{B}^{a} &=&\partial l^{a}=\frac{\sqrt{2}}{2}(0,\frac{1-\overline{\zeta }%
^{2}}{1+\zeta \overline{\zeta }},\frac{-i(1+\overline{\zeta }^{2})}{1+\zeta
\overline{\zeta }},\frac{2\overline{\zeta }}{1+\zeta \overline{\zeta }})
\nonumber \\
\overline{m}_{B}^{a} &=&\overline{\partial}l^{a}=\frac{\sqrt{2}}{2}(0,%
\frac{1-\zeta ^{2}}{1+\zeta \overline{\zeta }},\frac{i(1+\zeta ^{2})}{%
1+\zeta \overline{\zeta }},\frac{2\zeta }{1+\zeta \overline{\zeta }})
\nonumber \\
n_{B}^{a} &=&t^{a}-l^{a}=\frac{\sqrt{2}}{2}(1,-\frac{\zeta +\overline{\zeta }%
}{1+\zeta \overline{\zeta }},i\frac{\zeta -\overline{\zeta }}{1+\zeta
\overline{\zeta }},\frac{1-\zeta \overline{\zeta }}{1+\zeta \overline{\zeta }%
}).  \nonumber
\end{eqnarray}

We now introduce a (in general) twisting tetrad by performing an arbitrary
null rotation around $n_{B}^{a}$. Denoting the parameter of this rotation
(for historical reasons) by $-L$ , we obtain

\begin{eqnarray}
l^{a} &=&l_{B}^{a}-\frac{\bar{L}}{r}m_{B}^{a}-\frac{L}{r}\overline{m}%
_{B}^{a}+\frac{L\bar{L}}{r^{2}}n_{B}^{a}+O(r^{-3}),  \nonumber \\
m^{a} &=&m_{B}^{a}-\frac{L}{r}n_{B}^{a}+O(r^{-2}),  \nonumber \\
\overline{m}^{a} &=&\overline{m}_{B}^{a}-\frac{\bar{L}}{r}%
n_{B}^{a}+O(r^{-2}),  \nonumber \\
n^{a} &=&n_{B}^{a}.  \label{B-T}
\end{eqnarray}

Note that both tetrads are defined in the same Bondi coordinate system but
one of them, $l_{B}^{a},$ is surface forming while the twisted one is (in
general) not. For simplicity, we will refer to the tetrad system with the
`rotated' $l^{a}$ as the 'twisted tetrad' whether or not it has \textit{twist%
}. A special case, dealt with later, will be when $l^{a}$ \textit{is}
surface forming.

We now introduce an asymptotically flat smooth Maxwell field with tetrad
components by
\begin{eqnarray}
\phi _{0} &=&F_{ab}l^{a}m^{b}  \label{MaxField} \\
\phi _{1} &=&\frac{1}{2}F_{ab}(l^{a}n^{b}+m^{a}\overline{m}^{b})  \nonumber
\\
\phi _{2} &=&F_{ab}\overline{m}^{a}n^{b}.  \nonumber
\end{eqnarray}
In a Bondi frame their asymptotic behavior is given by
\begin{eqnarray}
\phi _{0B} &=&\frac{\phi _{0B}^{0}}{r^{3}}+O(r^{-4})  \label{BondiMax} \\
\phi _{1B} &=&\frac{\phi _{1B}^{0}}{r^{2}}+O(r^{-3})  \nonumber \\
\phi _{2B} &=&\frac{\phi _{2B}^{0}}{r}+O(r^{-2}),  \nonumber
\end{eqnarray}
the components satisfy the asymptotic Maxwell equations with Bondi
coordinates, $(u,\zeta ,\overline{\zeta }),$

\begin{eqnarray}
\left(\phi _{0B}^0\right)^\cdot+\text{$\frak{d}$}\phi _{1B}^{0} &=&0\;,  \label{lastMaxEqs} \\
\left(\phi _{1B}^0\right)^\cdot +\text{$\frak{d}$}\phi _{2B}^{0} &=&0.  \nonumber
\end{eqnarray}
where ``dot'' and \text{$\frak{d}$} are the partial derivative with respect to u and the ``eth'' operator on the
sphere respectively. It is straightforward to show that the relationship between the twisted and Bondi
components are given by

\begin{eqnarray}
\phi _{0}^{0} &=&\phi _{0B}^{0}-2L\;\phi _{1B}^{0}+L^{2}\;\phi _{2B}^{0},
\label{null rot} \\
\phi _{1}^{0} &=&\phi _{1B}^{0}-L\;\phi _{2B}^{0},  \nonumber \\
\phi _{2}^{0} &=&\phi _{2B}^{0}.  \nonumber
\end{eqnarray}

We then restrict the choice of this function $L$ by requiring that the
asymptotic shear associated with the $\ell ^{a}$ congruence vanishes at $%
\frak{I}$. It then follows\cite{Aronson}  from this restriction that $
L(u_{B},\zeta ,\overline{\zeta })$ satisfies the pde

\begin{equation}
\text{$\frak{d}$}L+L\dot{L}=0. \label{Hspace}
\end{equation}

Using the relationships, (\ref{null rot}), the Eqs.(\ref{lastMaxEqs}) can be
rewritten for the twisted tetrad as,

\begin{eqnarray}
\left(\phi _{0}^0\right)^\cdot +\text{$\frak{d}$}\phi _{1}^{0}+2\dot{L} \phi _{1}^{0}+L\left(\phi_{1}^0\right)^\cdot &=&0\;,  \label{TwistMaxEqs} \\
\left(\phi _{1}^0\right)^\cdot +\text{$\frak{d}$}\phi _{2}^{0}+\left(L\phi _{2}^{0}\right)^\cdot &=&0, \nonumber
\end{eqnarray}
where (\ref{Hspace}) has been used.

The solutions to (\ref{Hspace}) will play an important role in the
derivations given below. It is therefore instructive to review the method
used to obtain the solution together with the properties of the solutions.

We first introduce an auxiliary variable
\begin{equation}
\tau =T(u_{B},\zeta ,\overline{\zeta }),  \label{tau}
\end{equation}
which, by assumption, can be inverted to obtain
\begin{equation}
u_{B}=X(\tau ,\zeta ,\overline{\zeta }).  \label{cut}
\end{equation}
Making the ansatz that $L$ and $T$ are related by
\begin{equation}
L(u_{B},\zeta ,\overline{\zeta })=-\frac{\text{$\frak{d}$} T}{\dot{L}},  \label{L&T}
\end{equation}
it is straightforward to show, using implicit differentiation on (\ref{cut})
that

\[
L=-\frac{\text{$\frak{d}$} T}{\dot{L}}=\text{$\frak{d}$}_{(\tau )}X(\tau ,\zeta ,%
\overline{\zeta }),
\]
where $\text{$\frak{d}$}_{(\tau )}$ is the ``eth'' operator holding $\tau $ constant. Inserting the above in
(\ref{Hspace}) and using again implicit differentiation yields

\begin{equation}
\text{$\frak{d}$} L+L\;\dot{L}=\text{$\frak{d}$}_{(\tau )}^{2}X(\tau ,\zeta ,\overline{%
\zeta })=0.  \label{Hspace2}
\end{equation}

{\bf Remark 1}

An important point to be noted is that Eq.(\ref{L&T}) is unchanged by reparametrizing, with an arbitrary smooth
function $K,$
\begin{equation}
\tau \rightarrow \tau ^{*}=K(\tau )=K(T(u_{B},\zeta ,\overline{\zeta }%
))\equiv T^{*}(u_{B},\zeta ,\overline{\zeta }).  \label{reparam}
\end{equation}
This follows from
\[
L=-\frac{\text{$\frak{d}$} T^{*}}{\dot{T}^{*}}=-\frac{K,_{T}\,\text{$\frak{d}$} T}{K,_{T}\,%
\dot{T}}=-\frac{\text{$\frak{d}$} T}{\dot{T}}.
\]
This observation will be used later to simplify some expressions.

The general (regular) solution to (\ref{Hspace2}) is given by
\begin{eqnarray}
u_{B} &=&X(\tau ,\zeta ,\overline{\zeta })=\xi ^{a}(\tau )l_{a}(\zeta ,%
\overline{\zeta }),  \label{u3} \\
L(u_{B},\zeta ,\overline{\zeta }) &=&\xi ^{a}(T)m_{a}(\zeta ,\overline{\zeta
}),  \label{b3}
\end{eqnarray}
with $\xi ^{a}(\tau )$ four arbitrary complex analytic functions of the
complex $\tau .$ This is interpreted as a complex analytic world-line in
complex Minkowski space. Note that $\xi ^{a}(\tau )$ can be decomposed into
the two parts

\begin{equation}
\xi ^{a}(\tau )=\xi _{R}^{a}(\tau )+i\xi _{I}^{a}(\tau )  \label{world-line}
\end{equation}
where ($\xi _{R}^{a}(\tau ),$ $\xi _{I}^{a}(\tau )$) are both \textit{real
analytic} functions (i.e., their respective Taylor series coefficients are
real). In the special case of $\xi _{I}^{a}(\tau )=0,$ it turns out that the
twist, $\Sigma (u_{B},\zeta ,\overline{\zeta }),$ defined by
\begin{equation}
2i\Sigma =\text{$\frak{d}$}\bar{L}+L\;\dot{\bar{L}}-\bar{\text{$\frak{d}$}}L-\bar{L}\dot{L} \nonumber
\end{equation}
vanishes and the `twisted' tetrad is surface forming. In that case, Eq.(\ref{u3}) can be (correctly) interpreted
as the intersection from the future light cone of the real world-line $\xi _{R}^{a}(\tau _{R})$ with $\frak{I}$.
$\tau _{R}$ is the variable $\tau $ taken as real. It describes the family of cuts of $\frak{I}$ obtained by
following the light-cones from the world-line $\xi _{R}^{a}(\tau _{R})$. In the more general case, with the
complex world-line, the cuts of $\frak{I}$ from Eq.(\ref{u3}) are complex, so that their interpretation is via
the twisting real congruence.

In the general case with an arbitrary complex world-line, (\ref{world-line}), we can give a geometric interpretation of the function $L(u_{B},\zeta ,%
\overline{\zeta }).$ At any point of $\frak{I}$, ($u_{B},\zeta ,\overline{\zeta }$ ), the pair
($L,\overline{L}$) are stereographic angles on the sphere of the past null cone. They thus represent on
$\frak{I}$, a field of null vectors pointing inwards. $\frak{I}$, with the pair ($L,\overline{L}$), can be
viewed as an inverted pin-cushion or porcupine. In general they do not line-up normal to any slicing of
$\frak{I}$.

\textit{In the special case, when the twist}, $\Sigma =0,$ t\textit{hey do line-up and are orthogonal to a
slicing given by }$\tau =$ $real$ $const$. We can then introduce a second \textit{real} coordinate system in a
neighborhood of $\frak{I}$, the so-called NU system with its associated null tetrad. We recall that the Bondi
coordinates at $\frak{I}$ can be thought of as the intersection of the future light cone of a time-like geodesic
with $\frak{I}$. This is generalized to the time-like curve given by $\xi _{R}^{a}(\tau _{R})$. The new cuts of
$\frak{I}$, $\tau _{R}=const$ will be 'wiggly' cuts when written in terms of the Bondi coordinates. The explicit
coordinate relationship between these two sets of cuts is expressed via the (now real) relation
\begin{equation}
u_{B}=\xi _{R}^{a}(\tau _{R})l_{a}(\zeta ,\overline{\zeta })  \label{NU}
\end{equation}
or its inverse $\tau _{R}=T(u_{B},\zeta ,\overline{\zeta })$. The tetrad is
\textit{geometrically the same} `twisted' tetrad but now described in the ($%
\tau ,\zeta ,\overline{\zeta })$ coordinates. Performing the straightforward
but slightly lengthy coordinate transformation, (\ref{NU}) on (\ref
{TwistMaxEqs}), leads us to the set

\begin{eqnarray}
\text{ }V(\phi _{0}^{0})^{\prime }+\text{$\frak{d}$}_{(\tau )}(V^{2}\phi _{1}^{0})
&=&0  \label{MaxwellinNU} \\
(\phi _{1}^{0})^{\prime }+\text{$\frak{d}$}_{(\tau )}(V\,\phi _{2}^{0}) &=&0, \nonumber
\end{eqnarray}
with prime denoting the $\tau $-derivative and
\begin{equation}
V=(\xi _{R}^{a\,})^{\prime }\,l_{a}.  \label{V}
\end{equation}

{\bf Remark 2}

We want to point out that there is the very useful formal trick of allowing the introduction of complex
coordinates that allows considerable simplification in certain integrals. Instead of using Eq.(\ref{NU}) where
$\tau $ and $u_{B}$ are real, we can go back to Eq.(\ref{u3}),
\begin{equation}
u_{B}=\xi ^{a}(\tau )l_{a}(\zeta ,\overline{\zeta }),  \label{NU2}
\end{equation}
and treat it simply as a complex coordinate transformation between the complex $\tau $ and $u_{B}.$ The
relations given by (\ref{MaxwellinNU}) are unchanged in form and can be used with complex $\tau .$ We refer to
this complex coordinate system as \textit{complex} NU coordinates. Note that the Maxwell equations (with the
same tetrad system) are considerably simpler than their counterparts, Eq.(\ref{TwistMaxEqs}), in the real Bondi
coordinates.

Since some of our relations, e.g., Eq.(\ref{u3}), involve both real and
complex variables, it is worth a brief discussion this issue. We have
assumed from the beginning that we were dealing with a real analytic Maxwell
fields so that the real coordinates could, \textit{if needed, }be extended
into the complex. The complex conjugate stereographic pair ($\zeta ,%
\overline{\zeta })$ could be freed from each other, ($\zeta ,\overline{\zeta
})\rightarrow $($\zeta ,\widetilde{\zeta })$. Thus considering both $u$ and $%
\tau $ to be complex, makes Eq.(\ref{u3}) meaningful. Nevertheless we are
interested only in real values for $u$. Assuming, of course, the
invertibility of (\ref{u3}), we take the values of $u$ as real and find the
associated values of $\tau $ as ($\zeta ,\overline{\zeta })$ range over the
real sphere, thus for real, $u=const$, mapping the sphere into the complex $%
\tau $-plane. For a real cut, $u=const$, there are a spheres worth of points
in the complex Minkowski space so that for each point there is a null ray,
given by Eq.(\ref{b3}), that intersects the cut. Hence from a real point of
view the complex world-line becomes a world-tube, $S^{2}$x$R,$ in complex
Minkowski space. Every point on the tube corresponds to a real null
direction on $\frak{I}$. There is a degenerate situation when the tube
collapses into a real world-line that occurs when a real world-line is
chosen.

The question now is how to chose the complex world-line in a canonical
fashion so that it can be identified as the (complex) center charge world
line. The basic idea is to generalize the usual idea of the (real) center of
charge where there is a moving 'origin' about which there is no electric
dipole moment to the situation where there is a \textit{complex world-line}
about which there is \textit{neither an} electric $n$\textit{or} a magnetic
dipole moment. These moments are usually associated with the $l=1$ harmonic
coefficients of the component $\phi _{0}^{0}$ of the Maxwell field. Our task
is to determine $\xi ^{a}(\tau )$ and hence the null rotation to $l^{a},$ so
that this occurs for $\phi _{0}^{0}$.

If the Maxwell field in question has a \textit{principle null vector} that
is tangent to a shear free null geodesic congruence, then if we chose it as
part of our twisting tetrad system then automatically the full $\phi _{0}$
vanishes hence forcing $\phi _{0}^{0}=0.$ This then determines the $\xi
^{a}(\tau )$ uniquely. In the next section we consider a special case of
this, namely the Lienard-Wiechert fields. It is important to realize that
this is a very special case. \{For Complex Lienard-Wiechert fields see
reference \cite{ShearFreeMax}\}. In Section IV it will seen how the LW case
can be generalized to the arbitrary radiation fields.

\section{Lienard-Wiechert fields}

In this section we review the Lienard-Wiechert Maxwell fields by first
recalling that these fields are generated by an arbitrarily moving point
charge in Minkowski space, $x^{a}=\xi _{R}^{a}(\tau _{R})$. The goal of this
section is to study, in this simple case, its connection with the discussion
of the previous section. We will see that the given source world-line, $%
x^{a}=\xi _{R}^{a}(\tau _{R}),$ coincides with the center of charge defined
by requiring that the $l=1$ harmonic coefficients of the component $\phi
_{0}^{0}$ vanishes. We then go on to integrate the Maxwell equations, in the
form of (\ref{MaxwellinNU}), and see how this leads to a definition of the
electric dipole moment. (In this case the magnetic dipole moment vanishes.)

We first remember from (\ref{NU}), that we can use the coordinates $(\tau
,\zeta ,\overline{\zeta })$ and the associated tetrad with the Maxwell
equations in the form (\ref{MaxwellinNU}). An important property\cite
{ShearFreeMax} of the LW fields that it has one principal null direction
associated with the null cones from the world-line. This implies that $\phi
_{0}=0,$ so that our condition $\phi _{0}^{0}=0$ is immediately identically
satisfied and, indeed (as was to be expected), the source line coincides
with our center of charge world-line. When $\phi _{0}^{0}=0$ is used in Eqs.(%
\ref{MaxwellinNU}), it leads to the simplification

\begin{eqnarray}
\text{$\frak{d}$}_{(\tau )}(V^{2}\phi _{1}^{0}) &=&0  \label{LWMaxwell} \\
(\phi _{1}^{0})^{\prime }+\text{$\frak{d}$}_{(\tau )}(V\,\phi _{2}^{0}) &=&0 \nonumber
\end{eqnarray}
with
\begin{equation}
V=(\xi _{R}^{a\,})^{\prime }\,l_{a}.
\end{equation}

The first equation is integrated as
\begin{equation}
\phi _{1}^{0}=Q(\tau )V^{-2}  \label{phi10}
\end{equation}
so that the second becomes
\begin{equation}
\text{$\frak{d}$}_{(\tau )}\text{(}V\phi _{2}^{0})=\{Q(\tau )V^{-2}\}^{\prime }. \label{Max2*}
\end{equation}
When Eq.(\ref{Max2*}) is integrated over the sphere (at fixed $\tau $), using the properties of $ \partial$ ,
one has
\[
\{Q(\tau )\int V^{-2}d\Omega \}^{\prime }=0.
\]
With $v^{a}=\xi _{R}^{a\,\prime }$ chosen to be a unit vector (by the
reparametrization of $\tau $) the integral term is constant and we have
conservation of charge, i.e.,

\[
Q(\tau )=q=\text{constant.}
\]
The last equation is integrated as\cite{Aronson}
\begin{eqnarray*}
\phi _{2}^{0} &=&qV^{-1}\overline{\text{$\frak{d}$}}_{(\tau )}[\frac{V^{\prime }}{V%
}] \\
&=&q[\xi _{R}^{b\,\prime }l_{b}]^{-1}\overline{\text{$\frak{d}$}}_{(\tau )}[\frac{%
\xi _{R}^{a\,\prime \prime }l_{a}}{\xi _{R}^{c\,\prime }{}l_{c}}].
\end{eqnarray*}
The asymptotic 'shear-free' LW Maxwell field is then given, in the 'twisting
tetrad system', by
\begin{eqnarray}
\phi _{0} &=&0  \label{TheSolution} \\
\phi _{1} &=&qV^{-2}r^{-2}+...  \nonumber \\
\phi _{2} &=&qV^{-1}\overline{\text{$\frak{d}$}}_{(\tau )}[\frac{V^{\prime }}{V} ]r^{-1}+....  \nonumber
\end{eqnarray}

Note that in the particle's instantaneous rest-frame, where $V=1,$ we have
the standard dipole radiation term proportional to the second derivative of
the electric dipole moment, $q\xi _{R}^{a}(\tau )$.

\begin{equation}
\phi _{2}=\frac{q\xi _{R}^{a\prime \prime }\overline{m}_{a}}{r}+...
\label{elecdipolerad}
\end{equation}
with $q\xi _{R}^{a}(\tau )$ the \textit{intrinsic dipole moment}.

Our goal however is not to study the properties of the LW field but to see how to extract the standard
definition of the electric dipole moment in standard Bondi coordinates and tetrad system, from the Maxwell
components. In principle these are the coordinates used by observers at infinity with standard clocks. We have
seen that by adopting a NU coordinate system that follows the motion of the source in the LW case, the Maxwell
component, $\phi _{0}$ vanishes. As we said earlier, the leading term, $\phi _{0}^{0},$ that normally contains
the dipole moment in \textit{just} the $l=1$ spherical harmonic term is automatically vanishing. However, for a
Bondi frame that is adapted to a fixed \textit{time-like} geodesic, the deviation of the world-line of the
charged particle from the time-like geodesic should give us a non-vanishing electric dipole moment. Solving for
$\phi _{0B}^{0}$ from (\ref{B-T}) we obtain

\begin{eqnarray}
\phi _{0B}^{0}(u,\zeta ,\overline{\zeta }) &=&2L\;\phi _{1}^{0}+L^{2}\;\phi
_{2}^{0},  \label{dipole} \\
&=&2L\;qV^{-2}+L^{2}\;qV^{-1}\overline{\text{$\frak{d}$}}_{(\tau )}[\frac{%
V^{\prime }}{V}]  \nonumber \\
&=&2q\xi ^{a}m_{a}[\xi ^{b\prime }l_{b}]^{-2}\{1-\frac{1}{2}(\xi ^{a}m_{a})[\xi ^{a\,\prime \prime
}\overline{m}_{a}-(\xi ^{a\,\prime \prime }l_{a})(\xi ^{c\,\prime }{}\overline{m}_{c}](\xi ^{b\prime
}l_{b})^{-1}\} \nonumber
\end{eqnarray}

with
\begin{equation}
u=\xi ^{a}(\tau )l_{a}(\zeta ,\overline{\zeta }).\label{u}
\end{equation}

The standard dipole is extracted from Eq.(\ref{dipole}) by choosing, at fixed $u$, the $l=1$ harmonic
coefficient, i.e., by integration with the weighting function, $\overline{m}_{a}.$ The (complex) dipole moment,
up to a scale factor, is then
\begin{eqnarray*}
D_{a}(u) &=&qY_{\bot a}(u)=\frac{3}{2\pi }\int \phi _{0B}^{0}\overline{m}%
_{a}dS. \\
\int m_{a}\overline{m}_{b}dS &=&\frac{4\pi }{3}\delta _{ab}.
\end{eqnarray*}

The symbol $\bot $ indicates that the quantity was perpendicular to the time direction. $D_{a}$ is then obtain
by substituting in every term of (\ref{dipole}) $\tau$ by $T(u,\zeta,\bar{\zeta})$. To prepare ourselves for
this calculation we observe that
$$V \dot{T}=1,$$
(as can be easily shown from (\ref{u})) and rewrite $\phi _{0B}^{0}$ as
$$
\phi _{0B}^{0}=2q\xi ^{a}m_{a}\dot{T}^{2}\{1-\frac{1}{2}(\xi ^{a}m_{a})[\xi ^{a\,\prime \prime
}\overline{m}_{a}-(\xi ^{a\,\prime \prime }l_{a})(\xi ^{c\,\prime }{}\overline{m}_{c}](\xi ^{b\prime
}l_{b})^{-1}\}.
$$
Using the recursion formulas for products of spherical harmonics, in principle the $\ell=1$ can be extracted.
This, however, could be a highly non trivial task. If we consider slow motion of the sources we can approximate
the $D_{a}$ as follows: Choose, by reparametrization, that
\[
u=\tau +\xi ^{i}(\tau )l_{i}(\zeta ,\overline{\zeta })=\tau +\frac{1}{2}\xi
^{b}(\tau )(l_{b}-n_{b}),
\]
and then approximate its inversion by
\[
T =u-\frac{1}{2}\xi ^{b}(u)(l_{b}-n_{b}).
\]
We find, up to linear order in $\dot{\xi} ^{b}(u)$, that
$$
\dot{T} =1-\frac{1}{2}\dot{\xi}^{b}(u)(l_{b}-n_{b}),
$$
\[
\phi _{0B}^{0}(u,\zeta ,\overline{\zeta })=2q[\xi_d(u)-i\frac{\sqrt{2%
}}{4}\epsilon_{dac}\xi ^a\dot{\xi}^{c}]m^{d}+\{l\eqslantgtr 2,harmonics\}.
\]
In this calculation the Clebsh-Gordon expansion
\[
m_{a}[l_{b}-n_{b}]=\frac{i\sqrt{2}}{2}\epsilon _{abd}m^{d}+\{l=2,harmonics)\}
\]
was used.

We then have the conventional electric and magnetic dipole moments given in Bondi coordinates from the complex
$D^{d}(u),$ as

\begin{equation}
Y^{d}(u)m_{d}=\xi ^{d}(u)m_{d}-i\frac{\sqrt{2}}{4}\xi ^a (u)\dot{\xi}^{b}(u)t^{e}\epsilon _{eabd}m^{d}.
\label{Y1}
\end{equation}
The spacial vector $Y^{d}(u)$reads
\begin{equation}
Y^{d}(u)=\xi_{\bot }^{d}(u)-i\frac{\sqrt{2}}{4}\xi ^a (u)\dot{\xi}^{b}(u)t^{e}\epsilon _{eab}{}^{d}, \label{Y}
\end{equation}
or, using vectorial notation,
$$\overrightarrow{Y}=\overrightarrow{\xi}(u)-i\frac{\sqrt{2}}{4}\overrightarrow{\xi} \times
\dot{\overrightarrow{\xi}}. $$ The dipole moment of $\phi _{0B}^{0}$ associated with the radiation field
generated from a moving charge $q$ with world line $\xi ^{d}(\tau)$ is
$$ \overrightarrow{D}(u) = q\overrightarrow{Y}(u).$$

We see that the `intrinsic complex dipole', $q\xi ^{a}(\tau ),$ appears
rather altered or disguised. Instead of $q\xi ^{a}(u);$we have terms like $%
q\xi ^{a}(T[u_{B},\zeta ,\overline{\zeta }])$ with other complications. If
we do not have the inversion of $u_{B}=\xi ^{a}(\tau )l_{a}(\zeta ,\overline{%
\zeta })$ for $\tau =T(u_{B},\zeta ,\overline{\zeta }),$ it is difficult to see what is the exact expression for
the dipole in the Bondi frame. The point is that the ``standard'' dipole is determined from the 'intrinsic'
dipole or equivalently from the null direction field, $L(u_{B},\zeta ,%
\overline{\zeta }),$ by a well-defined procedure. In different Bondi frames
it will take on different values while the intrinsic one remains unchanged.
To a good approximation, if the acceleration and velocities are small then
we do obtain

\[
D^{a}(u)=q\xi _{\bot }^{a}(u).
\]
i.e., from the part of the world line orthogonal to the time direction.

If the world-line was given by the straight $\xi ^{a}(\tau )=\delta
_{0}^{a}\tau +\delta _{i}^{a}R^{i},$ (in one Lorentz frame) then

\[
D^{a}(u)=qR^{a}.
\]
In this case if $R^{i}$ were imaginary, the magnetic dipole of charged
Kerr-Maxwell field would be obtained.

In the next section this idea is generalized to arbitrary asymptotically
vanishing Maxwell fields.

\section{General Maxwell Fields}

Returning to the general case of asymptotically vanishing Maxwell fields, we reverse the procedure described
above in the LW case, i.e., we will start with a Maxwell field described in a Bondi coordinate and tetrad frame
and then find, by a null rotation at $\frak{I}$, the appropriate twisting tetrad system that can be used to
determine the complex center of charge world-line $\xi ^{c}(\tau )$.

The Bondi form of the asymptotic Maxwell fields, Eqs.(\ref{BondiMax}), can
be determined by direct integration, from Eq.(\ref{lastMaxEqs}),

\begin{eqnarray}
\left(\phi _{0B}^0\right)^\cdot+\text{$\frak{d}$}\phi _{1B}^{0} &=&0,  \label{lastMaxII}
\\
\left(\phi _{1B}^0\right)^\cdot+\text{$\frak{d}$}\phi _{2B}^{0} &=&0.  \nonumber
\end{eqnarray}
From the arbitrary dipole data, $-q\ddot{Y}^a(u)\overline{m}_{a},$ with $Y^{a}(u)$ a complex four-vector
function of the real $u$, we have
\begin{equation}
\phi _{2B}^{0}=-q\ddot{Y}^a(u)\overline{m}_{a}+H_{l>1}^{(-1)}(u,_{a}%
\zeta ,\overline{\zeta }).  \label{dipoledata}
\end{equation}
The term $H_{l>1}^{(-1)}$ is included to indicate that other multipole fields might be present. For the
integration of the Eqs.(\ref{lastMaxII}) they have no interest for us. Since we are interested only in the
dipole (i.e., $l=1$) term and as the equations are linear no effects can enter from $H_{l>1}^{(-1)}$ $,$ so we
could totally ignore it for the moment. However non-linearities do enter later when we find the complex center
of charge line and by ignoring it we are leaving out small high order corrections. Our final results then are
only approximate in principle.

By straightforward integration we have

\begin{eqnarray}
\phi _{0B}^{0} &=&2qY^{a}(u)m_{a}  \label{0} \\
\phi _{1B}^{0} &=&q-q\dot{Y}^a(u)[l_{a}-n_{a}]  \label{1} \\
\phi _{2B}^{0} &=&-q\ddot{Y}^a(u)\overline{m}_{a}.  \label{2}
\end{eqnarray}
The idea is now to go to the twisting tetrad version of the Maxwell field

\begin{eqnarray}
\phi _{0}^{0} &=&\phi _{0B}^{0}-2L\;\phi _{1B}^{0}+L^{2}\;\phi _{2B}^{0},
\label{PHIs} \\
\phi _{1}^{0} &=&\phi _{1B}^{0}-L\;\phi _{2B}^{0},  \nonumber \\
\phi _{2}^{0} &=&\phi _{2B}^{0}  \nonumber
\end{eqnarray}
with
\[
L=\xi ^{c}(\tau )m_{c}
\]
and then determine the $\xi ^{c}(\tau )$ by requiring that the $l=1$ part of $\phi _{0}^{0}$ vanish. By
substituting Eqs.(\ref{0}), (\ref{1}), (\ref{2}) into the first of Eqs.$(\ref{PHIs})$ we obtain
\begin{eqnarray}
\phi _{0}^{0} &=&2qY^{a}(u)m_{a}-2qL(1-Y^{a\cdot }(u)[l_{a}-n_{a}])\;-L^{2}%
\;qY^{a\cdot \cdot }(u)\overline{m}_{a}  \label{PHI00} \\
              &=&2q[Y^{a}(u)-\xi^{a}(\tau)+\frac{i\sqrt{2}}{2}\epsilon^a{}_{bc}\xi^{b}(\tau)%
\dot{Y}^b(u)]m_{a}-q(\xi^{c}(\tau)m_{c})^{2}\;\ddot{Y}^a(u)\overline{m}_{a}. \nonumber
\end{eqnarray}

The process of exactly extracting the $l=1$ part and setting it equal to zero (at constant $\tau $) is difficult
and, in general, impossible. There are several reasons for this. The first is that both $\tau $ and angular
terms are hidden in the dependence of $Y(u)$ on $u=\xi ^{a}(\tau )l_{a}(\zeta ,\overline{\zeta })$. In addition
there are many products of spin-weighted spherical functions, ($m_{a},\overline{m}_{a},\,l_{a}$), that must be
decomposed. Nevertheless it is quite possible to setup a straightforward method to yield approximate solutions,
if we replace $\tau$  by $T(u,\zeta,\bar{\zeta})$ in the above expression and consider slow motion of the world
line $\xi(\tau)$. In this approximation the $\ell=1$ part of the equation reads
\[
Y^{a}(u)m_a=[\xi^{a}(\tau)-\frac{i\sqrt{2}}{2}\epsilon^a{}_{bc}\xi^{b}(\tau)\dot{Y}^b(u)]m_a
\]
Inserting $T =u-\frac{1}{2}\xi ^{b}(u)(l_{b}-n_{b})$ in the expression and solving for$\xi(u)$ up to first order
in $\dot{Y}^b(u)$ gives
\begin{eqnarray}
\xi ^{c}(u)m_{c} &=&Y^{c}(u)m_{c}-\frac{i\sqrt{2}}{4}\dot{Y}^a(u)Y^{b}(u)t^{e}\epsilon _{eabd}\eta ^{cd}m_{c} \\
\xi ^{c}(u) &=&u \delta _{0}^{c}+Y^{c}(u)-\frac{i\sqrt{2}}{4}\dot{Y}^a(u)Y^{b}(u)t^{e}\epsilon _{eabd}\eta
^{cd}+....
\end{eqnarray}

Note that in this equation $u$ is just a parameter to label the world line. Up to this order, it can be replaced
by $\tau$ to obtain $\xi ^{c}(\tau)$. Note also that it can be inverted to find $Y^{c}(u)$ to second order if we
were given $\xi ^{c}(\tau )$ and, as expected, it gives eq. (\ref{Y}) for L-W fields.

We have thus found the relationship between the complex center of charge world-line and the ordinary (electric
and magnetic) complex dipole moments.

\section{Conclusions}

We have argued here that for any asymptotically vanishing Maxwell fields
with non-vanishing total charge, there is a unique geometric structure,
namely a null direction field at $\frak{I}$ that is given by and conversely
gives, a unique complex analytic world-line in complex Minkowski space. This
world-line can be identified as the complex center of charge and when
multiplied by the total charge it defines the intrinsic dipole moment. These
structures exist independent of the choices of coordinates to describe them.
The usual definitions of the multipoles, either from the source point of
view or from the asymptotic fields, depend on the coordinate choice, are
derived from the intrinsic structures by integrations over the null
direction fields.

The argument, though we believe is new, depends only on already well known
and understood physics. It involves simply the observation that shear-free
and asymptotically shear-free null congruences have some very beautiful
properties that have not been fully exploited.

We mention that though in Minkowski space, `asymptotically shear-free'
implies `shear-free', we have not used that fact. The significance of this
observation lies in the fact that the material that has been presented here
is easily extended to Maxwell fields in asymptotically flat space-times
where the null congruences are only asymptotically `asymptotically
shear-free' and not in general `shear-free'.

As a final point we mention that the ideas presented here can be generalized
to asymptotically flat Einstein and Einstein-Maxwell fields where the role
of the complex center of charge world-line gets played by a complex center
of mass world-line. The role of the two dipole moments are played by the
spin and orbital angular momenta. In the case of Einstein-Maxwell fields
there will be two separate complex world-lines - the complex center of
charge and center of mass - different in general. If however they coincide,
it appears from preliminary calculations that this leads to the Dirac value
of the gyromagnetic ratio. These results will be reported elsewhere.

\section{Acknowledgments}

This material is based upon work (partially) supported by the National
Science Foundation under Grant No. PHY-0244513. Any opinions, findings, and
conclusions or recommendations expressed in this material are those of the
authors and do not necessarily reflect the views of the National Science.
E.T.N. thanks the NSF for this support. C.N.K. would like to thank CONICET
for support.

\end{document}